\documentstyle[12pt,aaspp4]{article}

\begin{document}

\title{Quantitative Estimates of 
  Environmental Effects on the Star Formation Rate of Disk Galaxies in
  Clusters of Galaxies}
\author{Yutaka Fujita\footnote{Present address; Department of 
  Earth and Space Science,
  Faculty of Science,
  Osaka University,
  Machikaneyama-cho, 
  Toyonaka, Osaka, 560-0043, Japan. fujita@vega.ess.sci.osaka-u.ac.jp}}
\affil{Department of Physics, Tokyo Metropolitan University,\\ 
 Minami-Ohsawa 1-1, Hachioji, Tokyo 192-03, Japan}
\authoremail{yfujita@phys.metro-u.ac.jp}

\begin{abstract}
A simple model is constructed to evaluate the change of 
star formation rate of a disk galaxy due to environmental effects in  
clusters of galaxies. 
Three effects, (1) tidal force from the potential well of the cluster, 
(2) increase of external 
pressure when the galaxy plows into the intracluster medium, 
(3) high-speed encounters between galaxies, are
investigated. General analysis indicates that the
star formation rate increases 
significantly when the pressure of molecular clouds rises above
$\sim 3\times 10^5 \rm\; cm^{-3}\;K$ in $\sim 10^8$ yr. This is because the
pressure rise makes the 
destruction time of the majority of
molecular clouds in the galaxy less than $10^8$ yr.
The tidal force from the potential well of the cluster accelerates molecular
clouds in a disk galaxy infalling towards the cluster center. 
Thus, the kinetic pressure rises above $\sim 3\times 10^5 \rm\;
cm^{-3}\;K$. Before the galaxy reaches the cluster 
center, the star formation rate reaches a maximum. The peak is three to
four times
larger than the initial value. If this is the main mechanism of the
Butcher-Oemler effect, blue galaxies are expected to be located 
within $\sim 
300$ kpc from the center of the cluster. However this prediction is
inconsistent with the recent observations. 
The increase of external 
pressure when the galaxy plows into the intracluster medium does not
change star formation rate of a disk galaxy significantly. 
Thus, the increase of external 
pressure may not be
the main mechanism of the Butcher-Oemler effect. 
The velocity
perturbation induced by a single high-speed encounter between galaxies 
is too small to 
affect star formation rate 
of a disk galaxy, while successive high-speed encounters
(galaxy harassment) trigger star formation activity because of the 
accumulation of gas in the galaxy center. Therefore, the galaxy harassment
remains as the candidate for a mechanism of the Butcher-Oemler effect. 

\end{abstract}

\keywords{galaxies: clusters: general --- ISM: clouds --- stars: formation}

\section{Introduction}

Clusters of galaxies in the redshift range of $0.2-0.5$ often exhibit an
overabundance, relative to present-day clusters, of blue galaxies
(Butcher \& Oemler \markcite{bo1978}1978). 
This star formation activity is often called the Butcher-Oemler effect 
(BOE).
Several mechanisms responsible for the effect have been
suggested.

Tidal compression by the cluster potential 
increases the velocity dispersion of clouds
in a disk galaxy (Byrd \& Valtonen \markcite{bv1990}1990 ; Valluri 
\markcite{v1993}1993). 
Henriksen \& Byrd \markcite{hb1996}(1996) estimate the influence of 
tidal acceleration 
by the cluster potential on galaxies. They show that if the cluster
potential is centrally peaked, compression of galaxy disk is
significant. 
They find that the velocity perturbation due to the 
gravitational tidal forces 
are large enough for molecular clouds to give birth to stars.

Dressler \& Gunn \markcite{dg1983}(1983) suggest that a significant
fraction of galaxies in a cluster 
undergo a rapid rise in external pressure as they
plow into the hot intracluster medium. 
Evrard \markcite{e1991}(1991) numerically confirms this possibility.
He suggests that the
high-pressure leads to star formation activity.

Recently, Moore et al. \markcite{mkl1996}(1996) 
numerically show that
successive high speed encounters, ``galaxy harassment'', cause a
complete morphological transformation from disks to spheroidals. Since 
galaxy harassment is an extremely effective mechanism for fueling
central region of a galaxy, it may cause star formation activity.

However, the above studies do not derive the history of star formation
rate (SFR) quantitatively.
In this paper, a simple method is devised to evaluate the SFR of a disk
galaxy in a cluster in order to find out what mechanism drives
the BOE. 
This model 
is useful to predict color and spatial distributions of
the blue population in BOE clusters. 
The above three mechanisms, 
(1) tidal force from the potential well of the cluster, 
(2) increase of external pressure, 
and (3) high-speed encounters between galaxies are 
studied as candidates of the BOE. 
These mechanisms 
provide pressure variations of local interstellar medium
(ISM) in a disk 
galaxy because of increase of kinetic or 
thermal pressure of ISM, and gas inflow towards the galaxy 
center. As will be shown in detail, 
the time scales of the variations are $\sim 10^8$, which reflect
the dynamical time scales of a galaxy and central region of a cluster.
Star formation responds this pressure variation locally. 
Elmegreen \& Efremov
\markcite{ee1997}(1997) obtain a simple relation between 
star formation efficiency of molecular 
clouds of varying
mass and pressure. Their results indicate that if the
pressure of a cloud increases the order of a magnitude, 
the star formation efficiency increases and the lifetime decreases 
significantly. 
The relation does not care about the origin of
the pressure. Thus, it can be applied to the above three cases (1) - (3). 
Since the lifetime of clouds is $\lesssim 10^8$ yr (Elmegreen \& Efremov
\markcite{ee1997}1997), the pressure variations in the cases
(1) - (3) are expected to affect the evolution of all the molecular clouds 
and the SFR of a disk galaxy. This will be discussed in the following
sections. 

The plan of this paper is as follows. In \S II, a model of
molecular cloud evolution is described. In \S III, the pressure evolution
of a disk galaxy in a cluster is derived, and the SFR is
evaluated. Finally, \S IV is devoted to conclusions. In that section,
observational implications are commented. 

\section{Molecular Cloud Models}

Molecular clouds are divided according to their {\em initial} 
masses, that is, $M_{\rm
  min}=M_1< ...< M_i< ...<M_{\rm max}$, where the
lower and upper cutoffs are $M_{\rm
  min}=10^2\;\rm M_{\sun}$ and
$M_{\rm max}=10^{8.5}\;\rm M_{\sun}$, respectively, and their intervals
are chosen that $\log(M_{i+1}/M_i)= 0.01$.  
Referring to the total mass of clouds whose initial masses are
between $M_i$ and $M_{i+1}$ as $\Delta \tilde{M}_i(t)$, the rate of change is
\begin{equation}
  \label{eq:rate}
  \frac{d \Delta \tilde{M}_i(t)}{dt} 
  = \tilde{f}_i [S_{\star}(t)+S_{\rm mol}(t)] - \frac{\Delta 
    \tilde{M}_i(t)}{\tau(M_i,P)}\;\,
\end{equation}
where $\tilde{f}_i$ is the initial mass fraction of the molecular clouds 
whose initial masses are
between $M_i$ and $M_{i+1}$, 
$S_{\star}(t)$ is the gas ejection rate from stars, $S_{\rm mol}(t)$ is
the recycle rate of molecular gas, 
$\tau(M_i,P)$ is the 
destruction time of a molecular cloud with mass $M_i$ and pressure $P$, 
respectively.

Since the 
mass spectra of open clusters, globular clusters, and molecular
clouds follow $N(M) \propto M^{-\alpha}$, where $\alpha = 1.6 - 2$
(Harris \& Pudritz \markcite{hp1994}1994 ; Elmegreen \& Efremov
\markcite{ee1997}1997), the initial mass function of molecular clouds is
assumed to be $N(M) \propto M^{-2}$. Thus $\tilde{f}_i$ can be derived 
from this relation.

The gas ejection rate from stars is divided into two terms, 
$S_{\star}(t) = S_s(t) + S_l$.
The gas ejection rate from stars with small lifetime is given by
\begin{equation}
  \label{eq:sou}
  S_s(t) = \int^{m_u}_{m_l} \psi(t-t_m)R(m)\phi(m)dm \:,
\end{equation}
where $m$ is the stellar mass, 
$\psi(t)$ is the SFR, $\phi(m)$ is the IMF
expressed in the form of the mass fraction, 
$R(m)$ is the return mass fraction, and $t_m$ is the lifetime of stars
with mass $m$.
The slope of the IMF is taken to be 1.35 (Salpeter mass function). The upper
and lower mass limits, $m_u$ and $m_l$, are taken to be 
$50 M_{\sun}$ and $2.8 M_{\sun}$, respectively, although the lower
limit of the IMF is  $0.08 M_{\sun}$. The lifetime of stars
and the return mass fraction are adopted from those in Mihara \& Takahara 
\markcite{mt1994}(1994) and Maeder \markcite{m1992}(1992), respectively. 
Note that the lifetime of a star with $2.8 M_{\sun}$ is $4\times 10^8$
yr, which is identical with 
the end time of calculations in the later sections. 
The gas ejection rate from stars with large lifetime, $S_l$, 
will be specified later.

The recycle rate of molecular gas is 
\begin{equation}
  \label{eq:rec}
  S_{\rm mol}(t) = \sum_i [1-\epsilon(M_i,P)]\frac{\Delta 
    \tilde{M}_i(t)}{\tau(M_i,P)}\;\,
\end{equation}
where $\epsilon(M_i,P)$ is the star formation efficiency 
of a molecular cloud with mass $M_i$ and pressure $P$.
Under simple assumptions, Elmegreen \& Efremov
\markcite{ee1997}(1997) derive $\epsilon(M_i,P)$ and
$\tau(M_i,P)$ : their results are adopted in the following. 
Note that the pressure $P$ includes the kinetic pressure (Elmegreen
\markcite{e1989}1989 ; Elmegreen \& Efremov \markcite{ee1997}1997). 
In our Galaxy, the kinetic pressure is larger
than thermal pressure (Elmegreen
\markcite{e1989}1989 ; Elmegreen \& Efremov \markcite{ee1997}1997). 

The SFR is described by  
\begin{equation}
  \label{eq:sf}
  \psi(t) = \sum_i \epsilon(M_i,P)\frac{\Delta 
    \tilde{M}_i(t)}{\tau(M_i,P)}\;.
\end{equation}

Equation (\ref{eq:rate}) can be solved if $P$ is given. 
In the next section, $P$ is determined for several cases. Then the 
SFRs are calculated form Equations (\ref{eq:rate})
$-$ (\ref{eq:sf}).

\section{Evolution of SFR of 
 Disk Galaxies in a Clusters of Galaxies}

\subsection{General Analysis}
\label{sec:gen}

In this subsection, the evolution of pressure is
treated as a parameter to find the relation between the 
evolution of pressure and
the SFR of a disk galaxy. 

Three cases are considered, that is, the final value of $P$ is
$P_{f} = 10$, 
$100$, and $1000 \;\rm P_{\sun}$ 
(Figure 1(a)). The pressure is given in units of
$\rm P_{\sun} = 3\times
10^{4} \;cm^{-3} \; K$ which is the total pressure in the solar
neighborhood (Elmegreen \& Efremov \markcite{ee1997}1997).
The pressure $P$
increases for $t>0$. For $t<0$, the model galaxy is assumed to be at
$P = \;\rm 1 P_{\sun}$

The calculation starts at $t = -10^{9}$ yr. At that time, the total 
mass of molecular clouds in the model galaxy, $M_{\rm tot}$, is
$10^{9.3} \;\rm M_{\sun}$, and the mass spectrum of molecular clouds is
$\propto M^{-2}$. 
For $t<0$, ignoring 
Equation (\ref{eq:sou}), the source term
$S_{\star}$ in Equation (\ref{eq:rate}) is artificially fixed at $S_0 = 
6 \;\rm M_{\sun}\; yr^{-1}$
which is the SFR of our Galaxy (G\'{u}sten \& Mezger
\markcite{gm1983}1983). 
After the calculation starts, the mass spectrum is 
modified because
the destruction time of molecular clouds depends on their mass 
(Elmegreen \& Efremov \markcite{ee1997}1997). 
Until $t=0$, 
the destruction and formation rates of molecular clouds 
become identical to within 3\%.
At $t=0$, $M_{\rm tot}=
2.4\times 10^{9} \;\rm M_{\sun}$, and the mass spectrum of molecular clouds is
$\propto M^{-1.7}$. The total 
mass of molecular clouds is similar to that of our Galaxy ($\sim 2\times 
10^{9} \;\rm M_{\sun}$, Larson \markcite{s1987}1987).
Note that for $t>0$ 
the SFR history normalized by $S_0$ is independent of the
value of $S_0$. 
The gas ejection rate from stars with large lifetime is given by 
$S_l = S_0 - S_s(0)$ for $t>0$ and for fixed $P_0$.

The evolution of pressure, SFR, and the total mass of
molecular clouds  
are indicated in Figure 1. As is shown, the SFRs and the time at which it
become maximum depends on $P(t)$.
Taking the case where $P_f = 10\rm\; P_{\sun}$
as an example, the SFR 
can be explained as follows. The pressure $P$
reaches $10\rm\; P_{\sun}$ at $t = 10^8$ yr.
When $P = 10\rm\; P_{\sun}$, 
the mass of a molecular cloud whose destruction time is less than
$10^8$ yr is 
$\lesssim 10^{8} \rm\; M_{\sun}$ 
(see Figure 4 in Elmegreen \& Efremov 
\markcite{ee1997}1997). 
On the other hand, 
when $t=0$, the total mass of molecular
clouds whose individual mass is less than $10^{8}\rm\; M_{\sun}$ is 
$1.6\times 10^{9} \rm\; M_{\sun}$. The ratio, 
$1.6\times 10^{9} \rm\; M_{\sun}/10^8\;\rm yr =
16\;\rm M_{\sun}\; yr^{-1}$, nearly corresponds to the 
peak of the SFR (Figure 1(b)).
In the case where $P_f = 100 \rm\; P_{\sun}$, the
SFR reaches a maximum before 
$P$ becomes $100\rm\; 
P_{\sun}$ ($t= 2\times 10^8$ yr).
This is because all molecular clouds ($\leq 10^{8.5} \rm\;
M_{\sun}$) are affected by the pressure increase before $t= 2\times 10^8$
yr, since the destruction time of a molecular cloud with $M = 10^{8.5}\rm\; 
M_{\sun}$ is $\sim 10^8$ yr when $P=100\rm\; P_{\sun}$.
In fact, $M_{\rm tot}$ is significantly reduced when 
$t= 2\times 10^8$ (Figure 1(c)). Therefore, the peak of the SFR
is determined by the balance between the time scale of pressure
increase and the destruction time of clouds. Moreover, Figure 1(b) shows that
the evolution of SFR is not very sensitive 
to $P_f$, if it is larger than $100\rm\; P_{\sun}$ and if 
the time scale of pressure increase is $\sim 10^8$ yr.
This is because almost all of the clouds are destroyed regardless of $P_f$.

In summary, when pressure increases on a time scale of $10^8$ yr, 
the SFR rises significantly if $P_f\gtrsim 10 \rm\;
P_{\sun}$ but reaches the ceiling at $P_f \sim 100\rm\; P_{\sun}$. 
The above arguments should be true for all of the BOE
mechanisms treated in the following sections, 
because this model does not care about
the origin of pressure. 

\subsection{A Radially Infalling Galaxy}
\label{sec:infa}

A radially infalling disk galaxy is affected by tidal force from the cluster
potential well. Molecular clouds in the galaxy are accelerated. Thus the
pressure $P$ from random cloud motions increases.

Three types of 
mass profiles in the central region of the cluster are investigated;
\begin{equation}
  \label{eq:gl}
  M_{\rm cl}(<r) = 1.2 \times 10^{14}\;{\rm M_{\sun}}\;
                   \left[\ln\left[\frac{r}{r_{\rm GL}}
                    +\left(1+\frac{r^2}{r_{\rm GL}^2}\right)^{1/2}\right]
                   -\frac{r/r_{\rm GL}}{(1+r^2/r_{\rm GL}^2)^{1/2}}\right]\:,
\end{equation}
\begin{equation}
  \label{eq:king}
  M_{\rm cl}(<r) = 3.5 \times 10^{14}\;{\rm M_{\sun}}\;
                   \left[\ln\left[\frac{r}{r_{\rm King}}
                    +\left(1+\frac{r^2}{r_{\rm King}^2}\right)^{1/2}\right]
                   -\frac{r/r_{\rm King}}
                   {(1+r^2/r_{\rm King}^2)^{1/2}}\right]\:,
\end{equation}
\begin{equation}
  \label{eq:nfw}
  M_{\rm cl}(<r) = 3.0 \times 10^{15}\;{\rm M_{\sun}}\;
                   \left[\ln\left(1+\frac{r}{r_{\rm NFW}}\right)
                   -\frac{r/r_{\rm NFW}}{1+r/r_{\rm NFW}}\right]\:.
\end{equation}
Equations (\ref{eq:gl}) is 
the profile inferred from gravitational lensing constraints, 
Equations (\ref{eq:king}) is the so-called King model profile,
and Equations (\ref{eq:nfw}) is the profile obtained in 
numerical simulations done by Navarro, Frenk, \& White
(\markcite{nfw1995}1995, \markcite{nfw1996}1996).
The characteristic radii for each 
are $r_{\rm GL}=50$ kpc, $r_{\rm King}=200$ kpc, 
and $r_{\rm NFW}=900$ kpc, respectively. 
The first profile is the same
as that in Henriksen \& Byrd \markcite{hb1996}(1996). The mass
distributions are normalized so that the mass in side 500 kpc is $2.5
\times 10^{14} \;\rm M_{\sun}$ for all distributions. When the mass profile 
of the cluster
follows Equation (\ref{eq:nfw}), the profile of the intracluster gas can be
fitted well by (\ref{eq:king}) except for the normalization of mass. 
In that case, 
the ratio of  characteristic radii, $r_{\rm King}/r_{\rm NFW}$, 
is 0.22 (Makino,
Sasaki, \& Suto \markcite{mss1998}1998), which is the reason for taking
a radius of 900 kpc in Equation (\ref{eq:nfw}). 

In order to calculate the tidal acceleration and the 
velocity perturbation of molecular clouds, the 
impulse approximation is used as in Henriksen \& Byrd
\markcite{hb1996}(1996). Strictly speaking, this approximation may
not be good because the rotation period of a galaxy is comparable to the 
time-scale in which the galaxy crosses the central region of the
cluster. However, as will be noted, the results are not very sensitive to 
the velocity
perturbations for the models considered in this subsection.

The transverse and radial tidal accelerations are 
\begin{equation}
  \label{eq:trans}
  a_t = GM_{\rm cl}(<r)\frac{R}{[R^2+(r+R)^2]^{3/2}}\;,
\end{equation}
\begin{equation}
  \label{eq:rad}
  a_r = GM_{\rm cl}(<r)\left[\frac{1}{r^2}-\frac{1}{(r+R)^2}\right]\;,
\end{equation}
where $G$ is the gravitational constant and $R$ is the
radius of the galaxy ($=20$ kpc).
The velocity
perturbation can be estimated as
\begin{equation}
  \label{eq:vmol}
  V = \int_0^t a dt \;,
\end{equation}
where $a$ is the transverse or radial 
tidal acceleration. The kinetic pressure is given by
$P_{\rm tidal} = 
\rho_{\rm d} V^{2}$, where the average density of the galaxy disk is 
$\rho_{\rm d} = 1.67\times 10^{-24}\rm\; g\;cm^{-3}$. The evolution of
$a$ is calculated from the orbits of the galaxy in the potential well of 
the cluster. As an idealized case,
a radially infalling galaxy is investigated. The velocity of the galaxy,
$v_{\rm gal}$, 
can be obtained by solving the equation of motion. 
At $t=0$, the position of the galaxy is 350 kpc away 
from the center of the cluster.
From there the galaxy radially infalls towards the center. The
initial velocity is $1000 \rm\; km\; s^{-1}$. Figure 2 shows the orbits
of the galaxy. 

Figures 
3(a) and 4(a) show the evolution of the pressure, $P = {\rm P_{\sun}} + P_{\rm
tidal}$, for the three types of potentials (Equations \ref{eq:gl}, 
\ref{eq:king}, and \ref{eq:nfw}). The rates of pressure increase 
due to the radial tidal acceleration are somewhat 
larger than those due to the transverse
tidal acceleration. Since the galaxy rotates, and radial and transverse 
tidal acceleration
cancel, the pressure of a real galaxy 
is less than that shown
in Figure 4(a). Note that when the galaxy disk is perpendicular to the
orbit, Figure 3(a) indicates the real tidal acceleration because there
is no radial tidal acceleration. 

Using the pressure evolution, Equation   
(\ref{eq:rate}) is solved. The evolution of a model galaxy for $t<0$ is the
same as that in \S\ref{sec:gen}. The evolution of the SFRs
is shown in Figure 3(b) and 4(b). 
The SFRs become maximum before the galaxy reaches the center of the
cluster. At that time, the galaxy is located 
at $50\lesssim r \lesssim 200$ kpc (Figure 2). 
Since the time scales of the pressure
increase are $\sim 2\times 10^8$ yr and $P$ becomes larger then $100 \rm\; 
P_{\sun}$ for all models, the evolution of the SFRs
is similar among them (see \S\ref{sec:gen}). 
Thus, it is common among the models 
that the maximum SFR is three to four
times as much as the initial SFR.

The evolution of $M_{\rm tot}$ is shown in Figure 
3(c) and 4(c). 
During the passage through the central region of a cluster, $M_{\rm tot}$ 
decreases to $\lesssim 1/5$ of the initial value.

\subsection{Pressure from the Intracluster Medium}
\label{sec:enc}

Numerical simulations of rich cluster evolution show that the local static 
pressure 
around a significant fraction of galaxies experienced a rapid increase at 
high redshift  (Evrard \markcite{e1991}1991). Thus,
the change of SFR due to this pressure increase is
investigated. The evolution of the pressure is shown in Figure 5(a) ; for
$0<t<3\times 10^8$ yr, $P= 10^{2 t/\rm Gyr} \;\rm P_{\sun}$, and for 
$t>3\times 10^8$ yr, $P= 10^{0.6} \;\rm P_{\sun}$. This corresponds to the
most rapid increase of pressure in Figure 1 of Evrard
\markcite{e1991}(1991). The evolution of a model galaxy for $t<0$ is the
same as that in \S\ref{sec:gen}. 
Figure 5 also shows 
the evolution of SFR
and $M_{\rm tot}$. The SFR increases by at most 70\%. Thus, the increase of 
static pressure may not induce observable changes in color. 

\subsection{High-Speed Encounters}
\label{sec:hse}

Moore et al. \markcite{mkl1996}(1996) propose that multiple high speed
encounters between galaxies (galaxy harassment) change the structure of  
galaxies in clusters. Therefore, the effect of high-speed
encounters on the SFR of a disk galaxy is considered.   

In the first place, a single encounter is considered.
As an extreme case, a penetrating encounter is calculated. The derivation of
$P(t)$ is the same as that in \S\ref{sec:infa} except for the mass
distribution of a perturber.  

The mass distribution of a perturber is
\begin{equation}
  \label{eq:per}
  M_{\rm per}(<r) = 1.0 \times 10^{11}\;{\rm M_{\sun}}\;
                   \left[\ln\left[\frac{r}{r_{\rm per}}
                    +\left(1+\frac{r^2}{r_{\rm per}^2}\right)\right]^{1/2}
                   -\frac{r/r_{\rm per}}{(1+r^2/r_{\rm per}^2)^{1/2}}\right]\:,
\end{equation}
for $r<50$ kpc. The characteristic radius $r_{\rm per}$ is 5 kpc.
For $r>50$ kpc, $M_{\rm per}(<r) = 2.0\times 10^{11}{\rm\; M_{\sun}}$.
The evolution of a model galaxy for $t<0$ is also the
same as that in \S\ref{sec:gen}. At $t=0$, a model disk galaxy is located 
at $r=100$ kpc, and the relative velocity to the perturber 
is $1000\rm\; km \;s^{-1}$.
Because of the symmetry between the approach and recession phases of the
model galaxy, only the transverse tidal acceleration is calculated. 
The results are shown in Figure 5. The model galaxy passes the center of 
the perturber at $t\sim 10^8$ yr. Since the kinetic pressure does not change
significantly, the SFR rises by only 13\%.

In this argument, however, the ISM of the perturber is not taken into account.
If it has its own ISM
with density $\rho_p$, the ram pressure $\rho_p V^2$ acts on 
molecular clouds in the model galaxy.  For example, when 
$\rho_d = 1.67\times 10^{-24} \rm\; g\; cm^{-3}$ for $r<5$ kpc and
$\rho_d = 0$ for $r>5$ kpc, the SFR temporarily increases more than
ten times larger than the initial value. The formation of a ring or
tidal arms due to a close encounter will also affect the SFR. 
However, Moore et al. \markcite{mkl1996}(1996) indicate that such close
encounters ($<30$ kpc) are extremely rare in a cluster. 
Thus, a single high speed
encounter does not drive the BOE. 

On the other hand, galaxy harassment would induce star formation
activity in many galaxies in a cluster. Using the results of Moore,
Lake, and Katz \markcite{mlk1998}(1998) and Lake, Katz, \& Moore
\markcite{lkm1998}(1998), the change of SFR due to the harassment is
calculated. 

Galaxy harassment has a great influence on galaxies smaller than the
Galaxy. Thus, the initial conditions of a model galaxy are different 
from those considered above. At $t=0$, the SFR is 
$1.2 \;\rm M_{\sun}\; yr^{-1}$ and $M_{\rm tot} = 3.7\times 10^9 \rm
\; M_{\sun}$, which correspond to those of the model galaxy in Moore et
al. \markcite{mlk1998}(1998) and Lake et
al. \markcite{lkm1998}(1998). 
The initial value of pressure is
taken to be $P = 0.1 \rm\; P_{\sun}$ in order to balance the
cloud destruction rate with the formation rate.
Lake et al. \markcite{lkm1998}(1998) show that most of the gas of 
the harassed galaxy
is driven to the region within 1 kpc of the center within 3 Gyr. In
the extreme case, half of the mass is transfered in an interval of 100 
- 200 Myr. Assuming that 60\% of the gas is driven to the region within 1
kpc of the center and that the hight of the gas disk is 200 kpc, the
gas density is $2.4\times 10^{-22}\rm\; g\; cm^{-3}$. Since the
effective velocity is $\sim 130\;\rm km\; s^{-1}$ (Fig. 2 in Moore at al. 
\markcite{mkl1998}[1998]), the pressure of the central region of the 
harassed galaxy is $1.0 \times 10^4 \;\rm P_{\sun}$. Thus, the
evolution of the model galaxy is assumed to be $P(t) = 10^{10 (t/\rm
Gyr)-1}\;\rm  P_{\sun}$ (Figure 5a). 
Using this and Equation (\ref{eq:rate}), 
the evolution of the model galaxy
can be calculated. The results are shown in Figure 5(b) and 5(c). 
The history
of SFR shows that galaxy harassment can induce strong star formation
activity. 

\section{Conclusions}
\label{sec:conc}

In this paper, the influence of environmental effects on the 
star formation rate (SFR) of disk galaxies in clusters was quantitatively 
estimated. The constructed model treats the response of star formation 
to local pressure variance. Since it does not care about the origin
of pressure, it can be used to investigate various candidates 
for the Butcher-Oemler effect. 

General analysis indicates that the
SFR increases significantly when the pressure of molecular clouds rises above
$\sim 10 \rm\; P_{\sun}$ in $\sim 10^8$ yr but 
reaches the ceiling at $\sim 100\rm\; P_{\sun}$. This is because the
destruction time of most 
molecular clouds in the galaxy becomes $\lesssim 10^8$ yr.  

In the analysis of a disk galaxy tidally affected by the cluster,
it is found that the maximum SFR is around three or four 
times as much as that of a
field galaxy, and does not depend greatly 
on the mass profiles of the cluster
and the type of the tidal force. The SFR becomes maximum before the galaxy
reaches the center of the cluster. At that time, the distance from the
center is $50\lesssim r \lesssim 200$.
Fritze -v Alvensleben \& Gerhard \markcite{fg1994}(1994) 
have investigated the color evolution during
and after starbursts for many types of galaxy. Their results show that
the color of a galaxy significantly becomes bluer 
for $\sim 10^8$ yr 
after the burst if the gas mass fraction just before the burst 
is $\sim 10$ \%, the duration of 
the burst is $\sim 10^8$ yr, and most of the gas is consumed during the
burst.  
Therefore, if the tidally affected
galaxies are responsible for the Butcher-Oemler effect (BOE), 
the blue galaxies should be 
found $\lesssim 300$ kpc from the center of the cluster (Figure 2).
It is confirmed that 
even if the initial velocity of the model galaxy is 2000 $\rm km\;
s^{-1}$, which is larger than the virial velocity of the model cluster, 
this prediction does not significantly change.
However, recent observations are inconsistent with 
this prediction. In A2317, the blue
population decreases remarkably inside 300 kpc (Rakos, Odell, \&
Schombert \markcite{ros1997}1997). Since this trend is observed in other
distant clusters (Abraham et al. \markcite{ash1996}1996 ; Balogh et
al. \markcite{bmy1997}1997), it appears that 
tidal force from the potential well of a
cluster does not induce the BOE. 

As a galaxy plows into the hot intracluster medium, the 
external pressure rapidly rises. However, the rise of local pressure
based on numerical calculation done by Evrard \markcite{e1991}(1991) does not
appear to induce starbursts because the maximum of the pressure is small.
This is consistent with the recent observation of a merging cluster
(Tomita et al. \markcite{tnt1996}1996).

The velocity
perturbation induced by a single high-speed encounter is too small to 
affect the SFR of a disk galaxy.   
However, successive high-speed 
encounters between galaxies (galaxy harassment) lead to gas inflow
and strong star formation activity.  
Thus, galaxy harassment may be a likely
explanation for the star-forming galaxies observed in high-redshift clusters
(Oemler, Dressler, \& Butcher \markcite{odb1997}1997 ; Couch et al. 
\markcite{cbs1998}1998 ; Smail et al. \markcite{see1998}1998). 
In order to know
whether galaxy harassment is the {\em main} origin of BOE, it is
required to estimate the 
fraction of galaxies in which the gas accumulates in the center. 
The spatial distribution of blue galaxies predicted by the galaxy
harassment model may be consistent with the observations, because
galaxies in the core of clusters will be older than galaxies at the 
edges. 

On the other hand, 
the BOE may not be excess star formation induced by the infall process
or internal tides in clusters of galaxies. Abraham et
al. \markcite{ash1996}(1996) suggest that the BOE is due to the
increased rate of infall of bluer field galaxies at higher redshift, and 
that the star formation is truncated without an increase. A model including
ram-pressure stripping and evolution of intracluster medium 
will be needed to investigate this idea.

\acknowledgments

I would like to thank S. Sasaki and T. Tsuchiya for their helpful comments. 
I wish to thank S. Inoue for his assistance. I am also grateful to an
anonymous referee for several suggestions that improved this paper. 
This work was supported in part by the JSPS Research Fellowship for
Young Scientists.

\newpage

\section*{Figure Captions}

Fig.1 --- Histories of (a) pressure (b) star formation rate, and (c)
total mass of molecular clouds.

Fig.2 ---- Orbits of radially infalling galaxies in the three potential
wells. Solid line; lensing potential (GL), dotted line; 
King model (King), dashed line; the one derived by Navarro et al.(NFW).

Fig.3 ---- Histories of (a) pressure (b) star formation rate, and (c)
total mass of molecular clouds for a galaxy affected by the transverse
tidal acceleration. Solid line; lensing potential (GL), dotted line; 
King model (King), dashed line; the one derived by Navarro et al.(NFW).

Fig.4 ---- Same as in Fig.3 but for radial tidal acceleration.

Fig.5 ---- Histories of (a) pressure (b) star formation rate, and (c)
total mass of molecular clouds for a galaxy affected by the pressure of 
intracluster 
medium (ICM; solid line), a single high-speed encounter (HE; dotted line), 
and galaxy harassment (GH; dashed line).

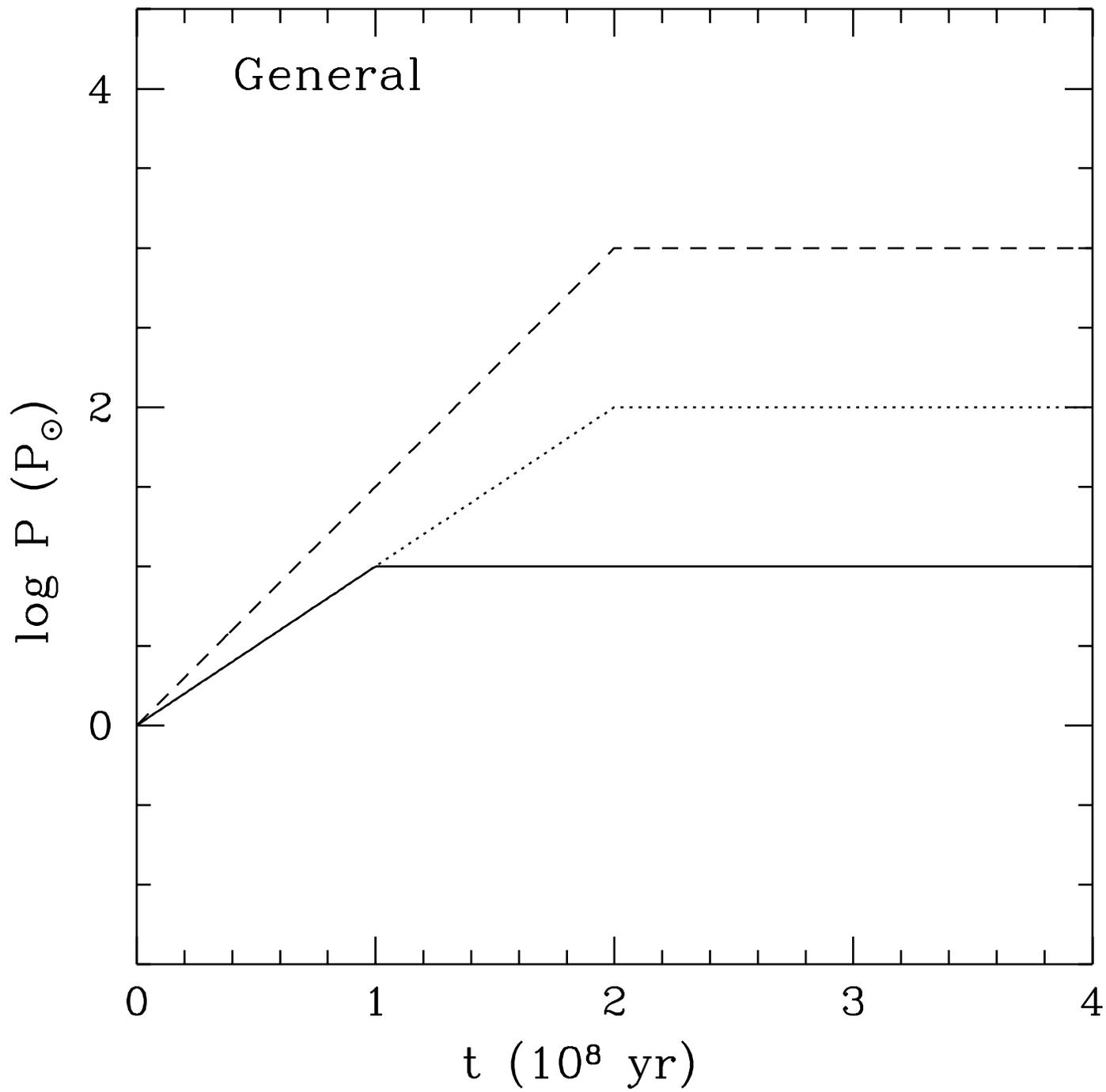
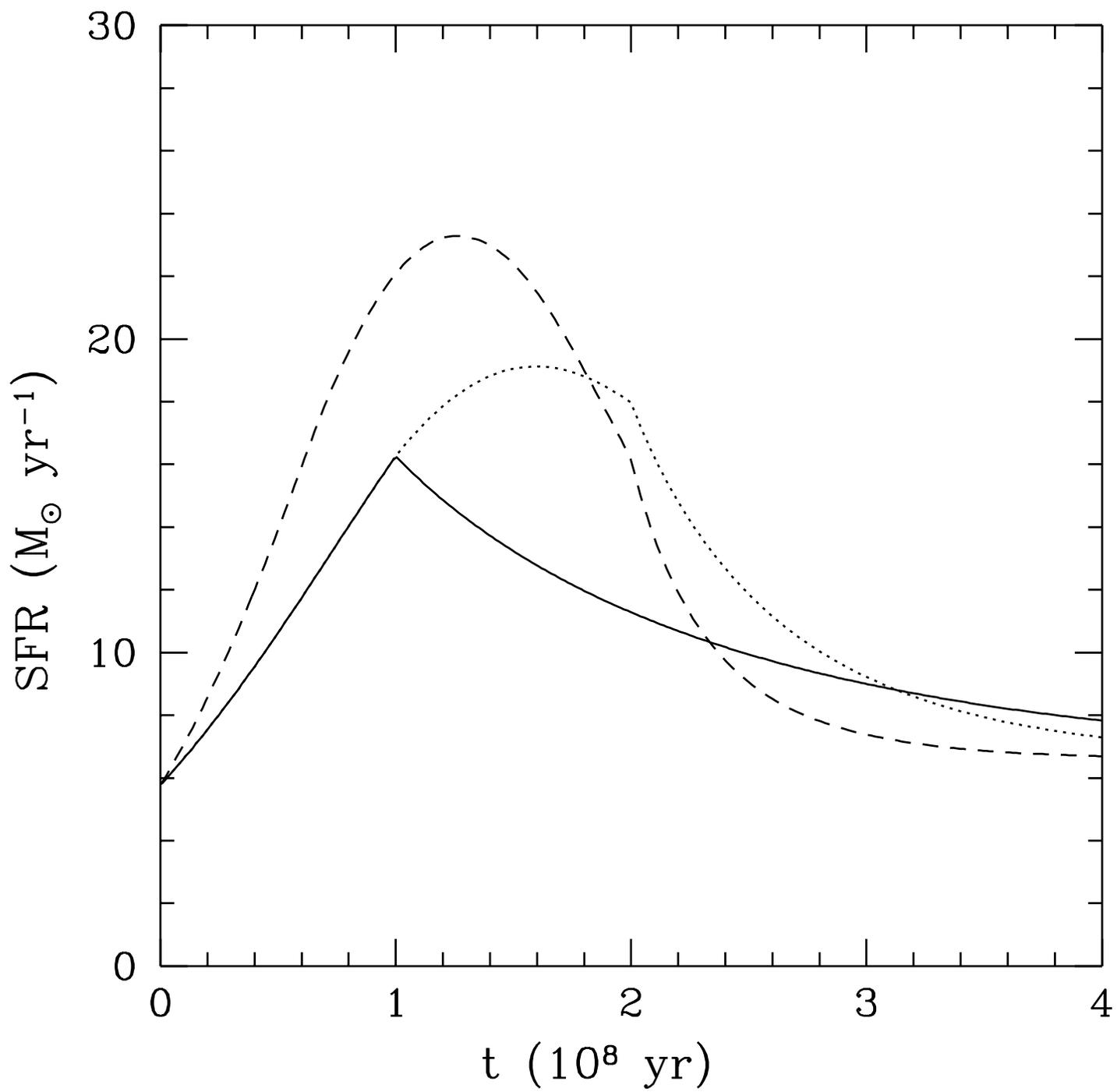
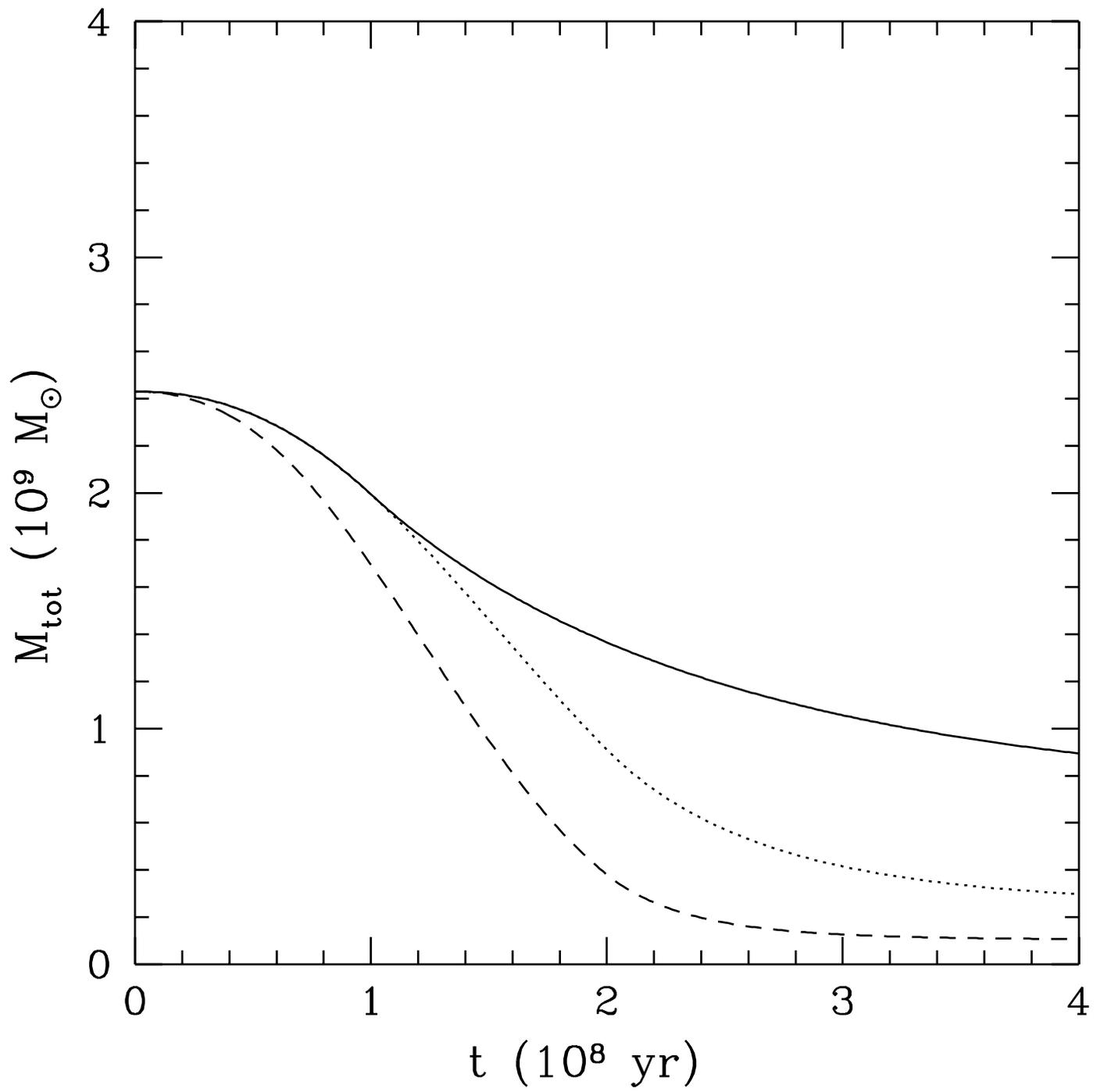
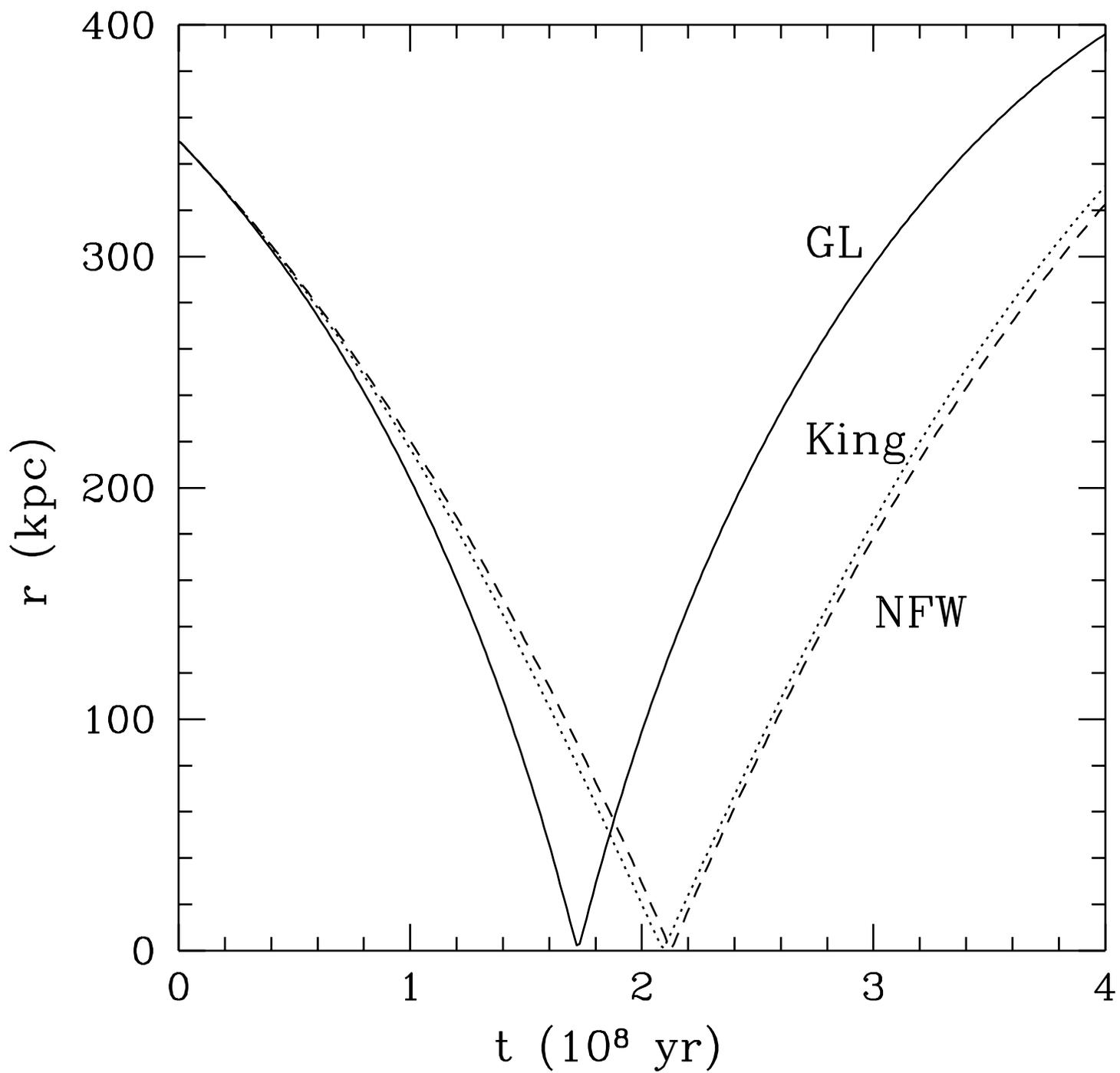
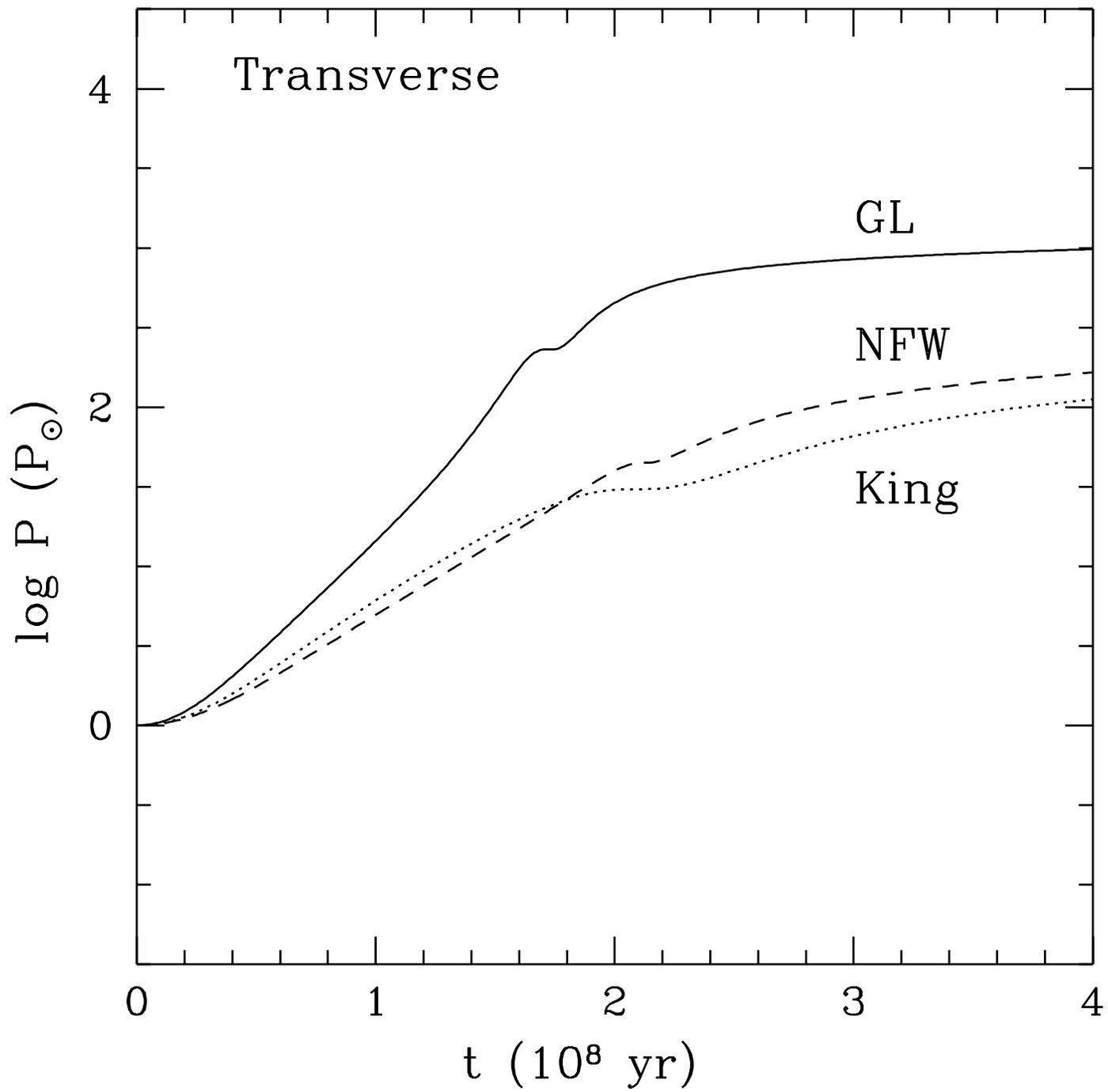
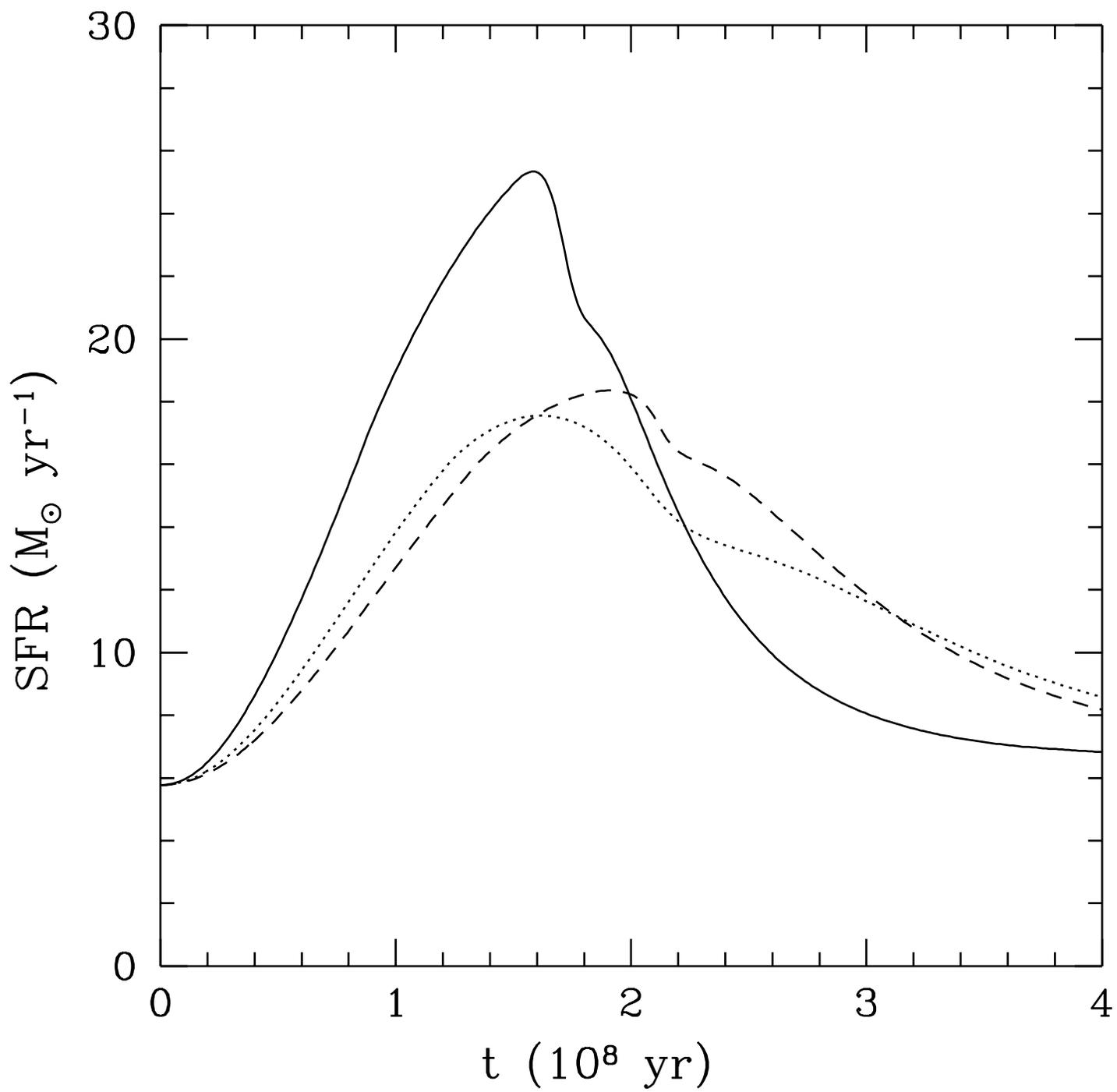
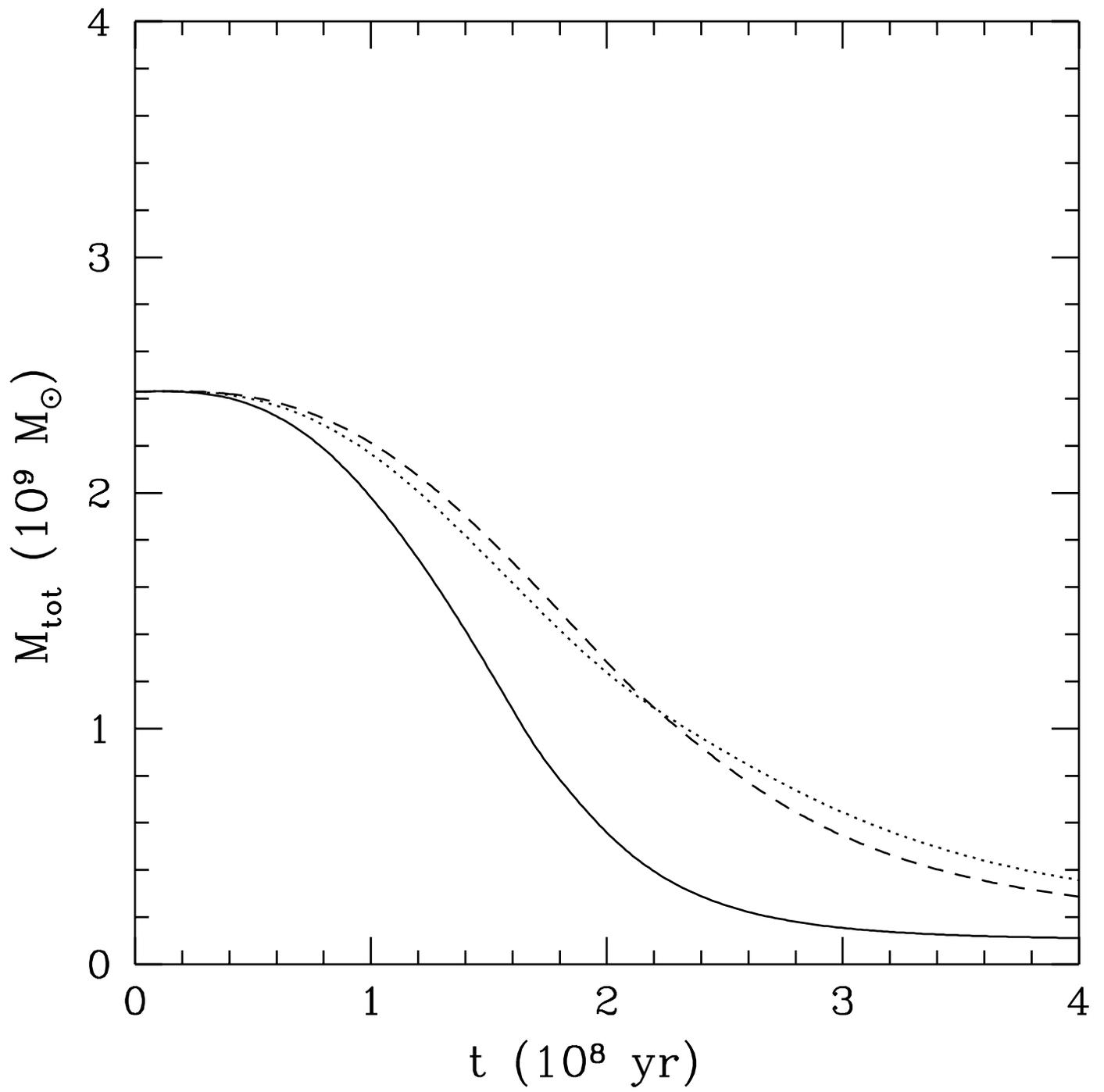
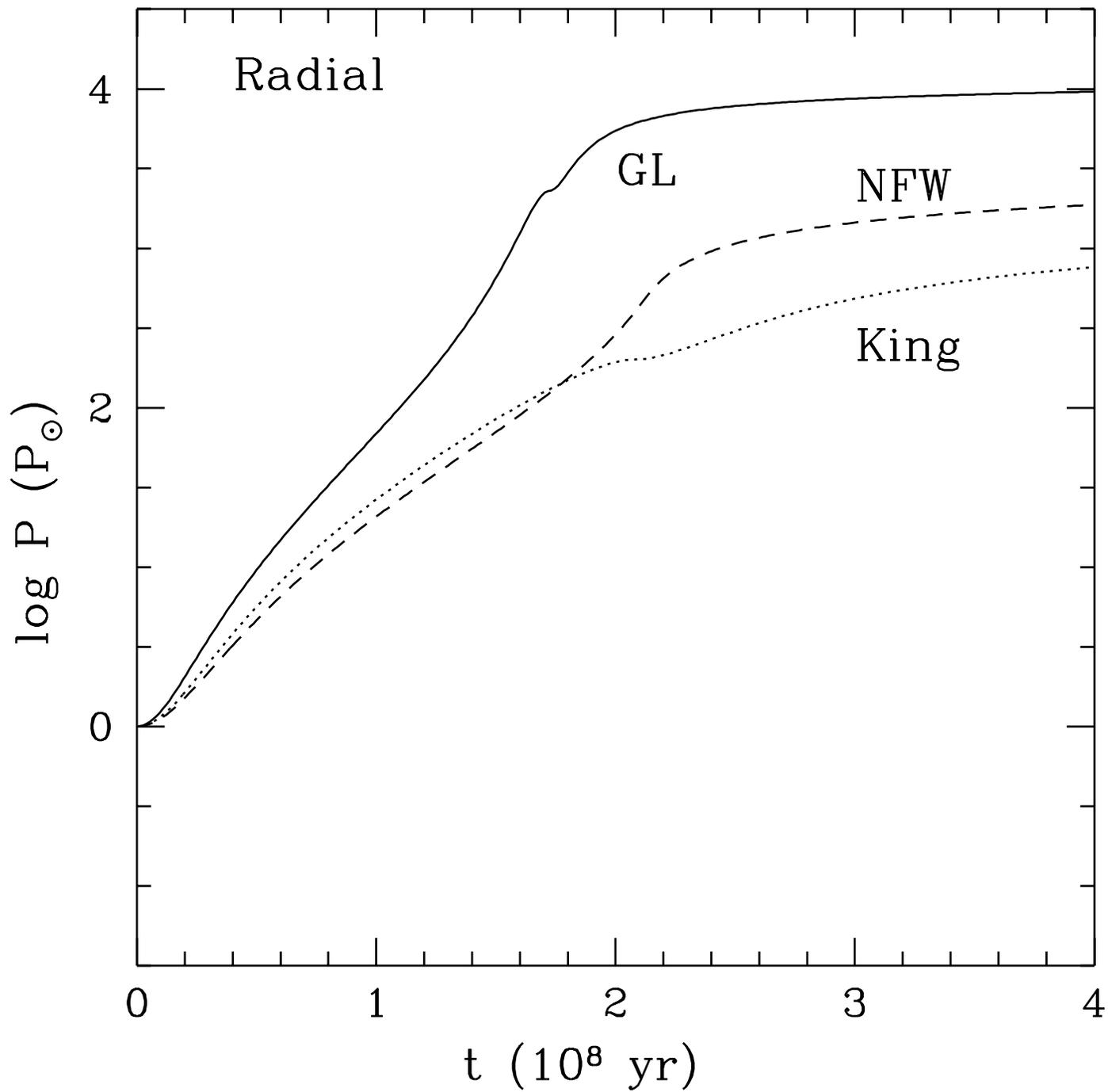
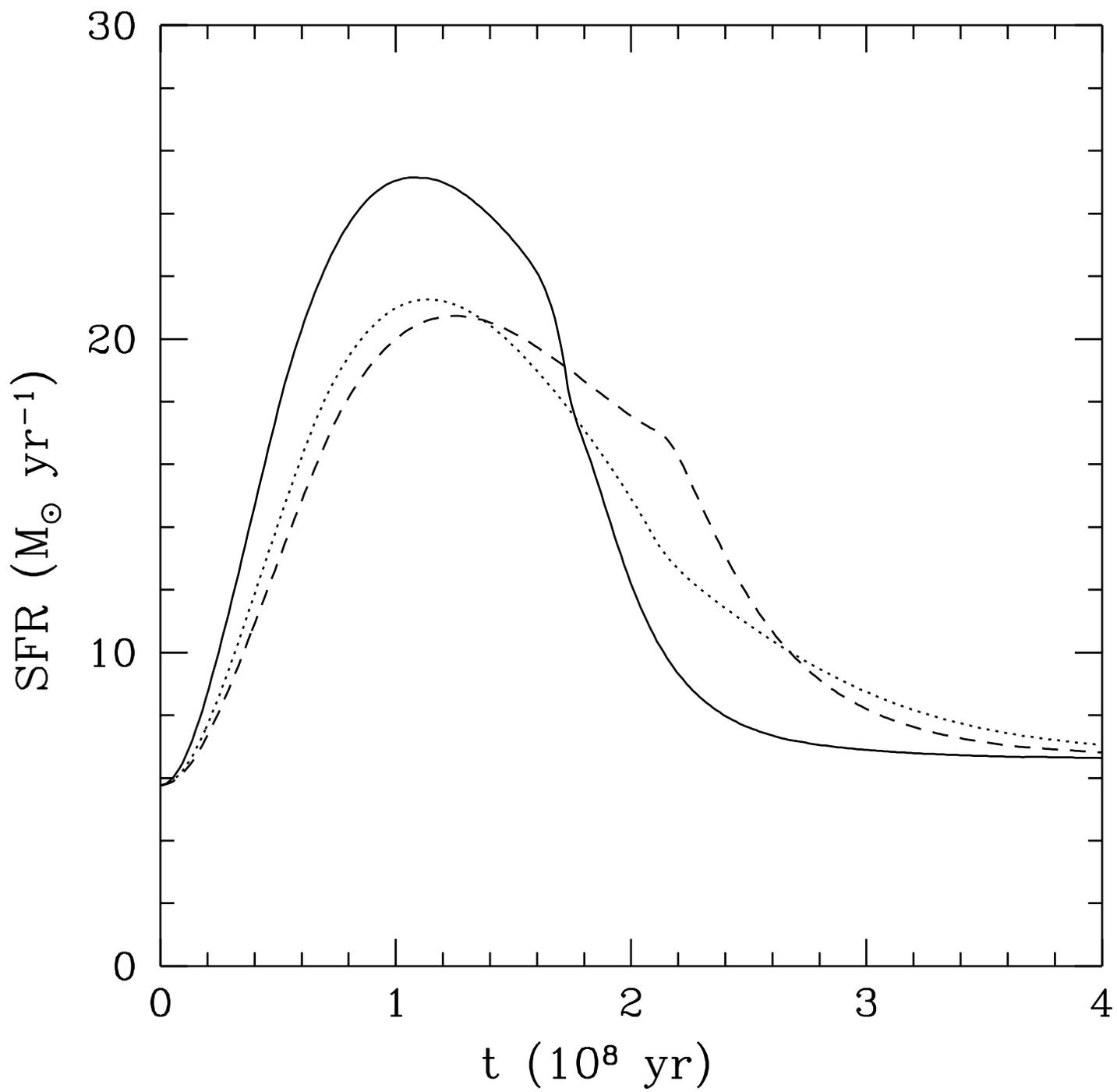
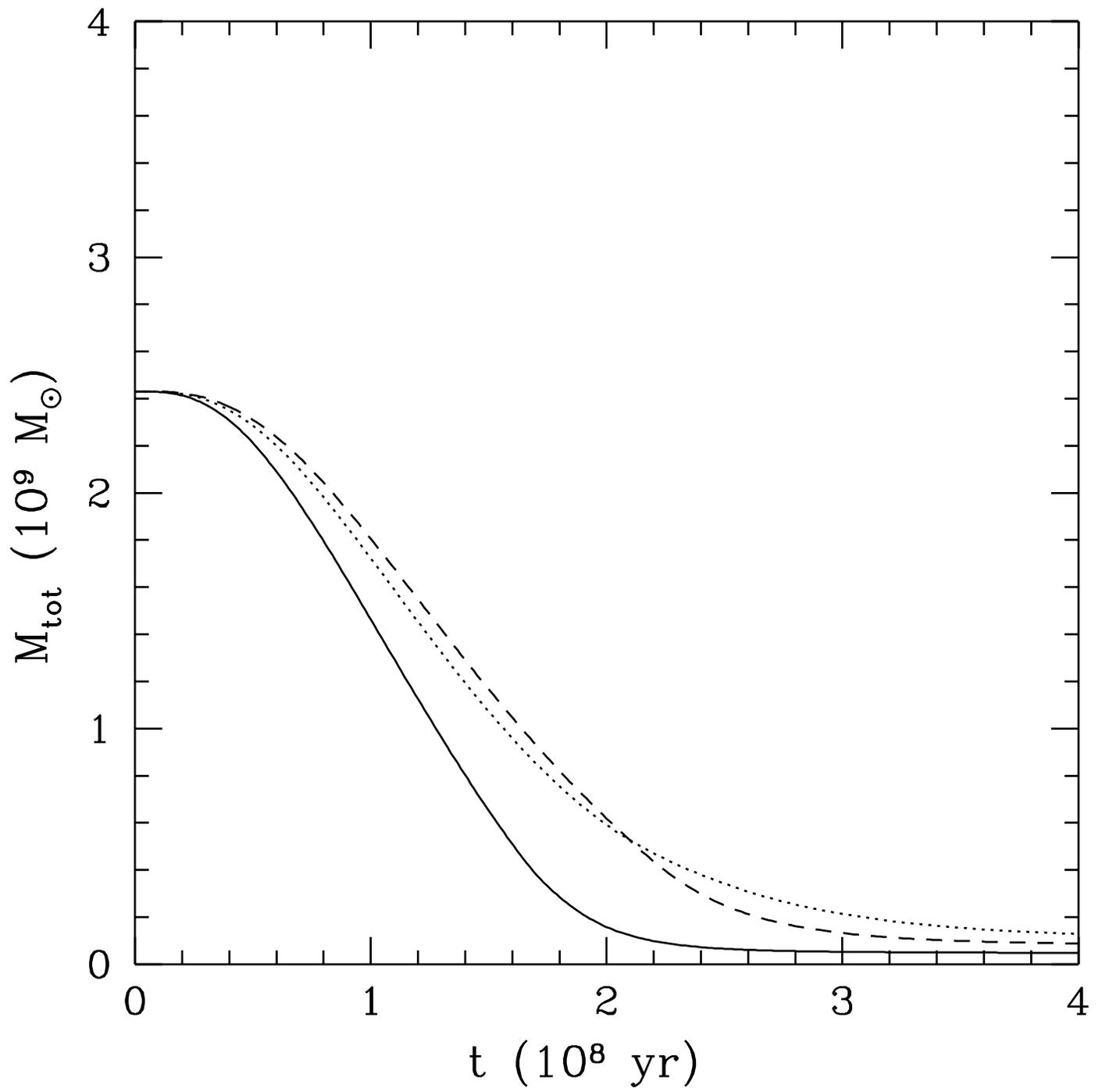
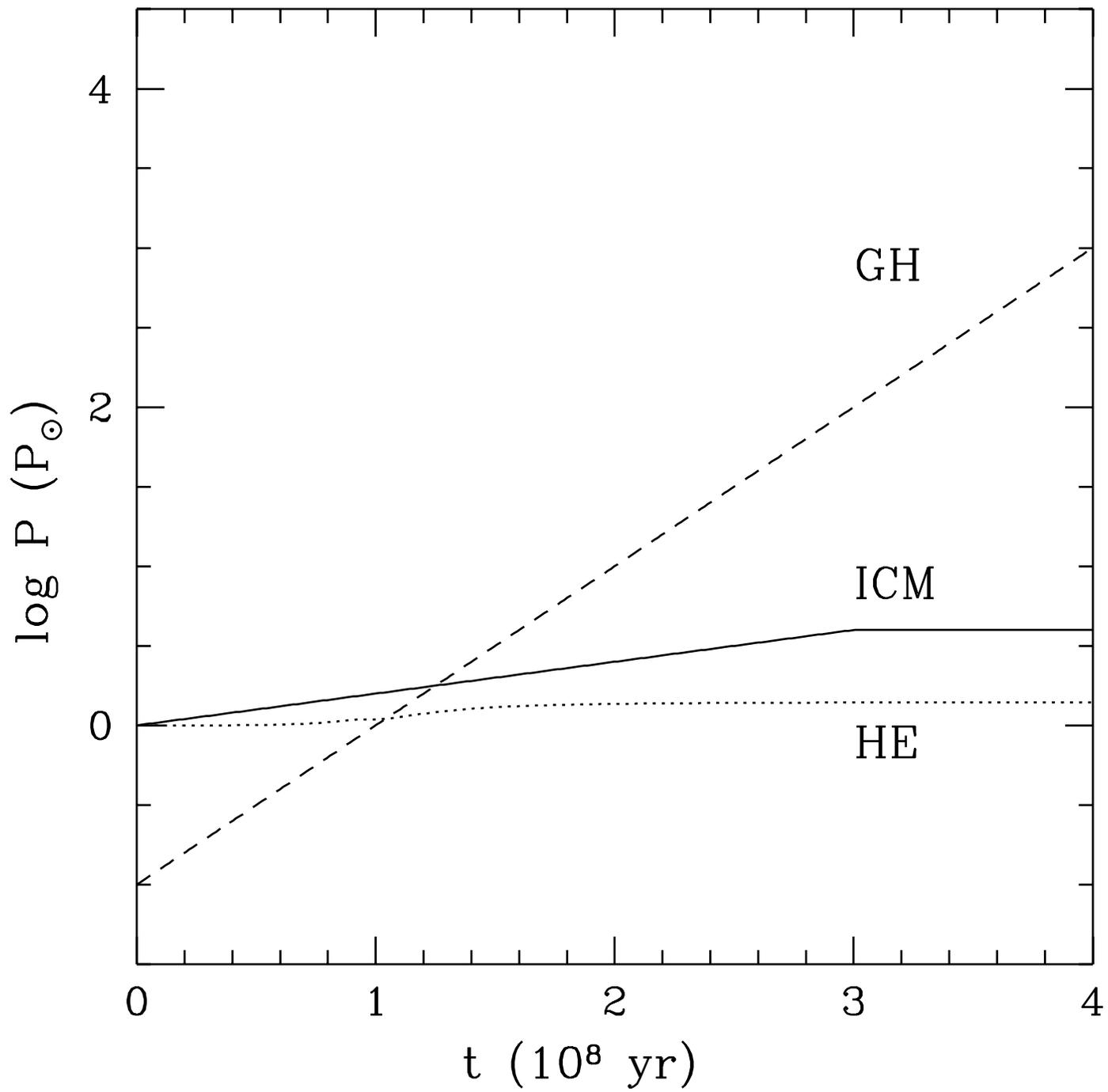
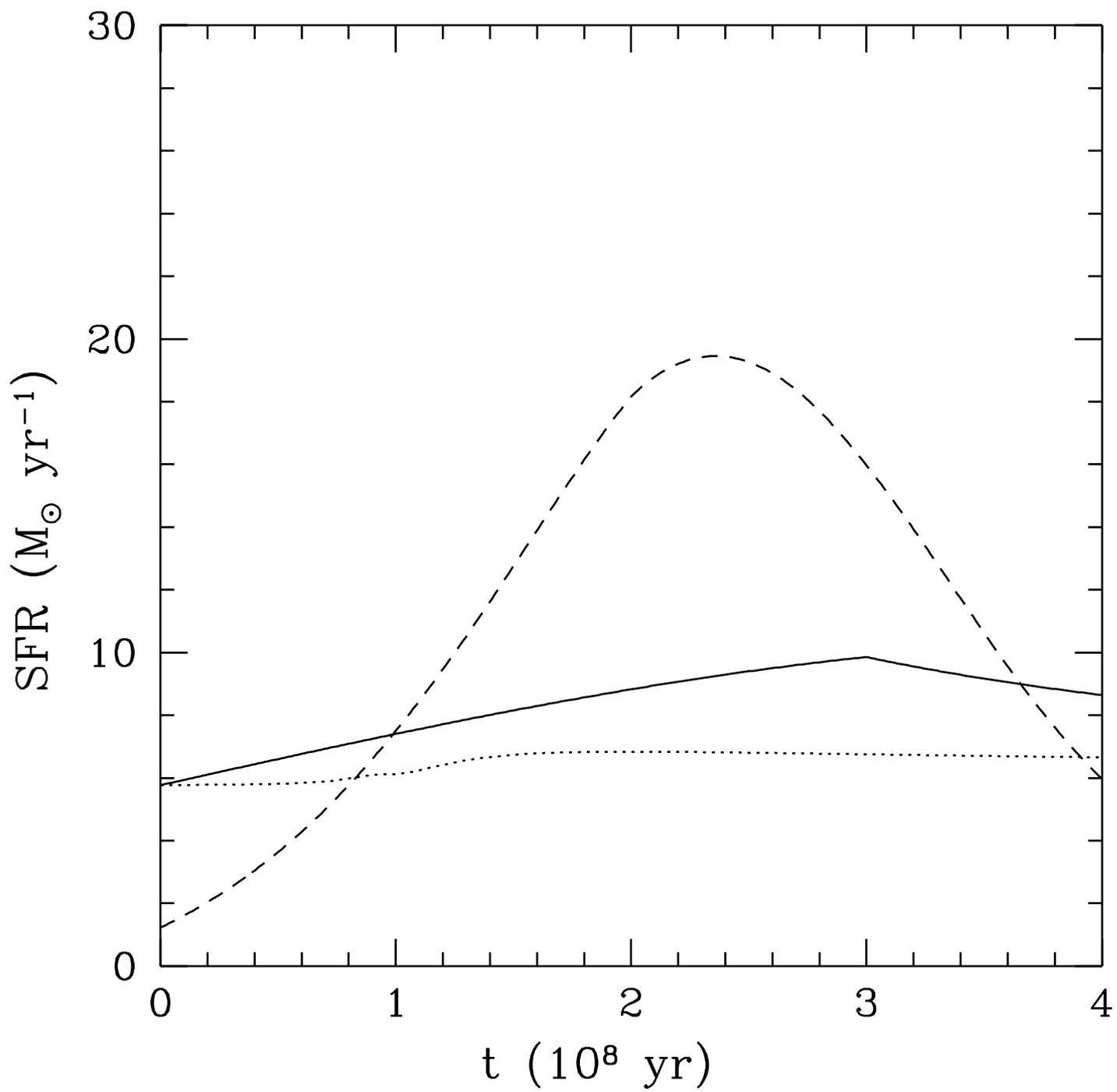
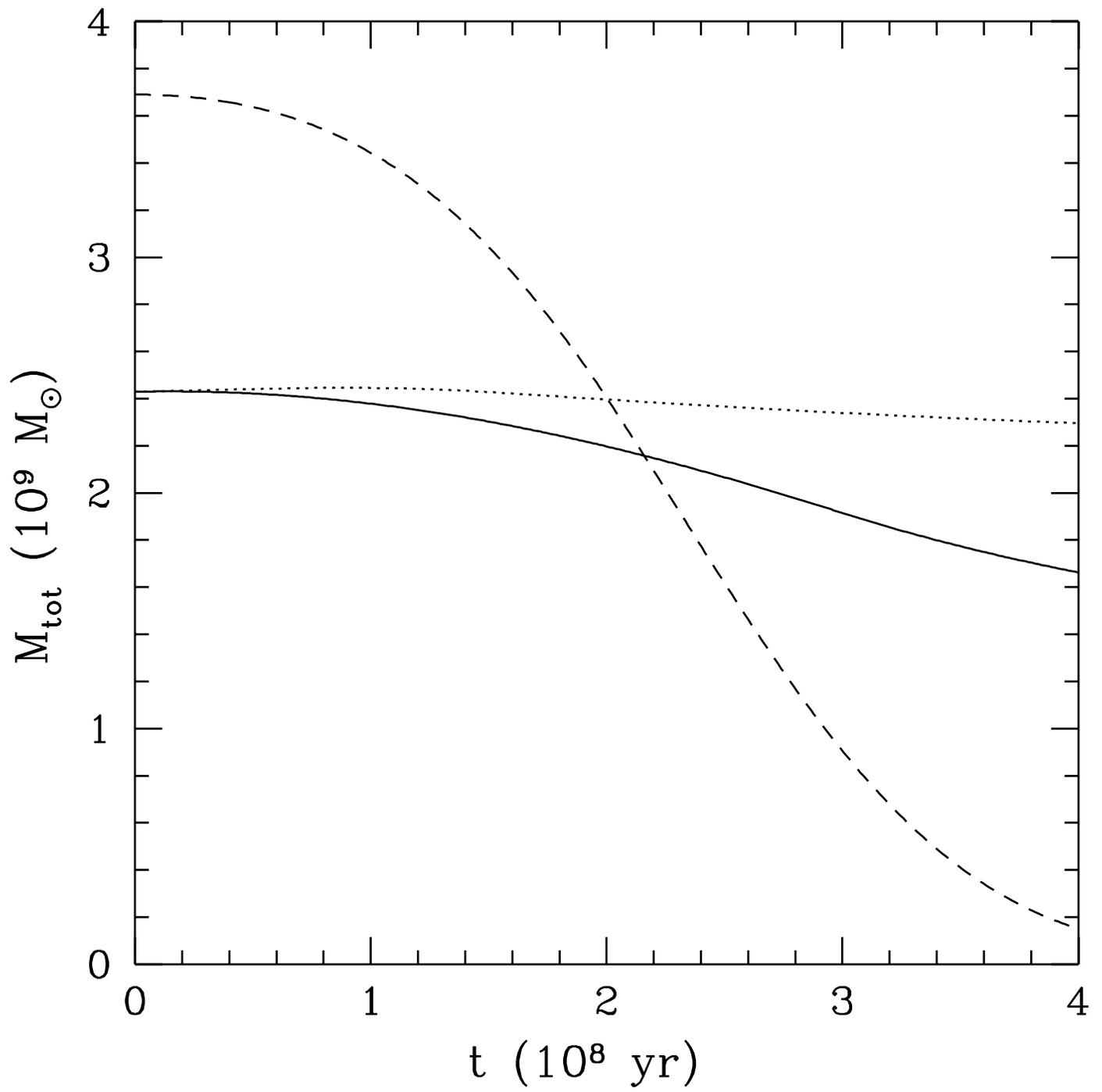
\end{document}